\documentclass[onecolumn,superscriptaddress,prb,preprint]{revtex4-1}
\usepackage{graphicx}
\usepackage{amssymb}
\usepackage{amsmath}
\usepackage{epsfig}
\usepackage{color}
\usepackage{mathtools}
\usepackage[colorlinks,linkcolor=blue,anchorcolor=blue,citecolor=blue,urlcolor=blue]{hyperref}

\begin{document}

\title{Super-resolution quantum sensing using NV centers based on rotating
linear polarized light and Monte-Carlo method}
\author{Hua-Yu Zhang}
\author{Xiang-Dong Chen}
\email{xdch@ustc.edu.cn}
\author{Guang-Can Guo}
\author{Fang-Wen Sun}
\affiliation{CAS Key Lab of Quantum Information, University of Science and Technology of China, Hefei,
230026, P.R. China}
\affiliation{Synergetic Innovation Center of Quantum Information $\&$ Quantum Physics, University of Science
and Technology of China, Hefei, 230026, P.R. China}
\date{\today}
\begin{abstract}
The nitrogen vacancy (NV) center in diamond has been widely applied for quantum information and sensing in last decade. Based on the laser polarization dependent excitation of fluorescence emission, we propose a super-resolution microscopy of NV center. A series of wide field images of NV centers are taken with different polarizations of the linear polarized excitation laser. The fluorescence intensity of NV center is changed with the relative angle between excitation laser polarization and the orientation of NV center dipole. The images pumped by different excitation laser polarizations are analyzed with Monte Carlo method. Then the symmetry axis and position of NV center are obtained with sub-diffraction resolution.
\end{abstract}

\maketitle


\section{Introduction}

With stable photon-emission and long spin coherence time at room
temperature, the negatively charged nitrogen-vacancy (NV) center in diamond
is a potential nanoscale quantum sensor. It has been applied in
the detection of electro-magnetic field \cite{lukin-nature2008-1,wra-nn-mag2015,Jacques-science2014-domainwall}, temperature \cite{wra-nl2013-them,chang-2015nl-tem,lukin-nature-therm},
and pressure \cite{Englund2016strain} with high sensitivity. Specially, NV center in nanodiamond is suitable for biological cell tracking \cite{changhc-2016acr} and sensing\cite{biosen,mag2} due to its small size and low cytotoxicity. On the other hand, various far field
super-resolution microscopies, such as stochastic optical reconstruction
microscopy (STORM) \cite{wra2014PNASstorm}, stimulated emission depletion (STED)
nanoscopy \cite{adma2012sil,pengxi-2014-rcs}, charge state depletion (CSD) nanoscopy \cite{hell-nl-micro,chen201501} and
quantum statistical image method \cite{cui-prl}, have been developed for NV
centers to achieve nanoscale spatial resolution. Combining with the super-resolution far-field microscopy, the quantum sensing and biological imaging with NV center can be realized with high spatial resolution.

For the quantum sensing with NV center, such as the detection of vector magnetic field, the external field induced resonant frequency shift of NV
highly would depend on the angle between the external field and the information of NV center symmetry axis. Therefore, it is necessary to
obtain orientation of each NV centers axes primarily. Although the
information of axes orientation can be achieved through optical \cite{wra-2014alignment,gumin2014polarization,Raymond2007PhysRevB-polarization} and
magnetic \cite{mag1} methods, the spatial resolution is still within the
diffraction limit. Here we present a super-resolution microscopy protocol
based on polarization demodulation \cite{polarization,dipole}. By scanning
the polarization of excitation laser, the photon intensity containing the
information of axes and positions of each NV centers are collected through a
CCD camera, and then the axes and positions of NV centers are estimated with Monte Carlo method.
Subsequently, the NV center can be located with sub-diffraction-limit
resolution. The orientation of symmetry axes and photon counts of each NV centers
can be obtained at the same time. Since the signal of quantum sensing processes
with NV center is usually readout through the photon counts, such a polarization
demodulation microscopy can be further applied in the wide-field
multi-functional quantum sensing with high spatial resolution.

\section{Method to get position and orientation}
\begin{figure}[h]
\centering\includegraphics[width=7cm]{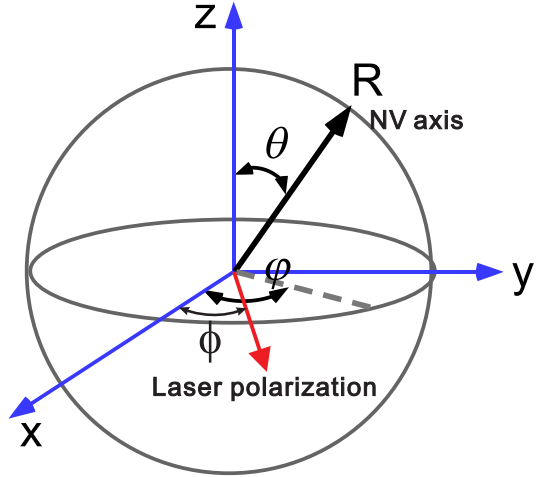}
\caption{Schematic drawing showing the relation of the NV center orientation and laser polarization. The orientation of NV axis is presented by a vector $R$ (with polar angle $\theta$ and azimuthal angle $\varphi$), and the polarization of laser on focusing plane is depicted by an angle $\phi$.}
\label{Fig1}
\end{figure}

When excited with linear polarized light, the photon intensity of each
NV center differs according to the relative angle between the NV
center symmetry axis and the laser polarization\cite{wra-2014alignment,jacques-2014alignment}. As shown in Fig.\ref{Fig1}, the photon intensity is related to $%
\theta $ and $\varphi $, which are the spherical angle of the orientation of
the NV center axis ($R$). And the excitation laser beam propagates along the $z$ axis. The fluorescence intensity of NV pumped by laser with polarization $\phi$ can be written as\cite{Raymond2007PhysRevB-polarization}:
\begin{equation}
I=A(1-\sin ^{2}\theta \cos ^{2}(\varphi-\phi))\text{,}  \label{E1}
\end{equation}%
where $A$ is the maximum photon counts and can be assumed as a constant for
 NV centers. In the experiment, the polarization (corresponding to $\phi
$) of the excitation laser is scanned from $0$ to $\pi $, and the NV center fluorescence emissions
(corresponding to the photon intensity $I_{Meas}(x,y,\varphi, \theta, \phi)$) are
recorded by a CCD camera. Here, we consider the point spread function (PSF) of
each NV center as a 2D Gaussian distribution. For the simulation of imaging process, we
assume the position of each NV center is discrete and the NV center is
located at the center of the CCD pixel. We define the energy ($E$) of the
system as the absolution of the difference between the measured fluorescence intensity distribution
($I_{Meas}(x,y,\varphi ,\theta, \phi )$) and the estimated fluorescence intensity distribution ($%
I_{Est}(x,y,\varphi ,\theta, \phi )$), which can be written as
\begin{equation}
E=\sum_{\phi=0}^{\pi}\sum_{x,y}Abs\left[ I_{Meas}(x,y,\varphi ,\theta, \phi
)-I_{Est}(x,y,\varphi ,\theta, \phi )\right]\text{.}
\end{equation}%
The estimated NV center positions and orientations those make the energy $E$
the smallest is the best fit answer.

Firstly, the locations and orientations of NV centers can be roughly estimated from the CCD images. The total number of NV centers in the image is roughly estimated from the total photon counts in the region. According to Eq.(\ref{E1}), for large number of NV centers, the
average fluorescence intensity of single NV is $0.75A$. So the number of NV centers in the region can
be estimated by dividing the total fluorescence intensity by a factor of $0.75A$. The local maximums in the images are set as the estimated positions of NV centers. The polarization dependence of fluorescence intensity at fixed CCD pixels are used to estimate orientation of the NV centers. For example, the polarization of laser changes with an initial polarization angle of 0 degree and a step of $\pi/{M}$. A total of  $M$ frames will be taken during the imaging process. If the weakest luminance appears in the $(k+1)$th frame, then
the estimated azimuthal angle of the NV center is $k\pi /M$. According to Eq.(\ref{E1}), the polar angle $\theta$ of NV center can be obtained as:
\begin{equation}
\theta =\arccos \sqrt{\frac{I_{min}}{I_{max}}}\text{,}
\end{equation}
where $I_{min}$ and $I_{max}$ are the measured minimum and maximum fluorescence intensities of each NV center pumped by laser with different polarizations, respectively.

A method called Metropolis sampling is used to further minimize the system
energy $E$. In the optimizing process, we randomly choose a single NV center from the
roughly estimated list, and change the estimated position or orientation angle a little. Then we
calculate the energy after the change. If the energy becomes smaller, we will
accept the new result. However, if the energy becomes larger, then
the new result will be accepted with a possibility $r=\exp (E_{1}-E_{2})$ obtained by the
Metropolis sampling method, where $E_{1}$ is the energy before the change
and $E_{2}$ is the energy after the change. After enough estimation steps the best fit of
position and orientation of each NV center can be obtained when $E$ is
convergent. After the
estimation processes is finished with a convergent energy $E$, we will
add or subtract an NV center in the high-density area to obtain a smaller
convergent energy $E$. The number of NV centers which makes the total energy
minimum is the best fit result.

\section{Simulated results}

\begin{figure}[h]
\centering\includegraphics[width=9cm]{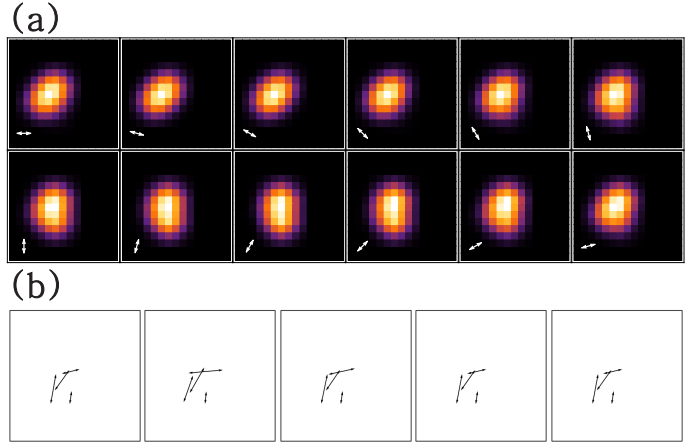}
\caption{(a) Simulated CCD image of 4 NV centers under linear polarized
laser pumping. The background noise is not considered here. The white arrows show the polarization of pump laser. (b) The first image on the left shows the real positions and orientations of NV centers. The next four images are the estimated positions and orientations of 4 NV centers after 200,
400, 600 and 800 estimation steps, respectively.}
\label{Fig2}
\end{figure}

\begin{figure}[h]
\centering\includegraphics[width=9cm]{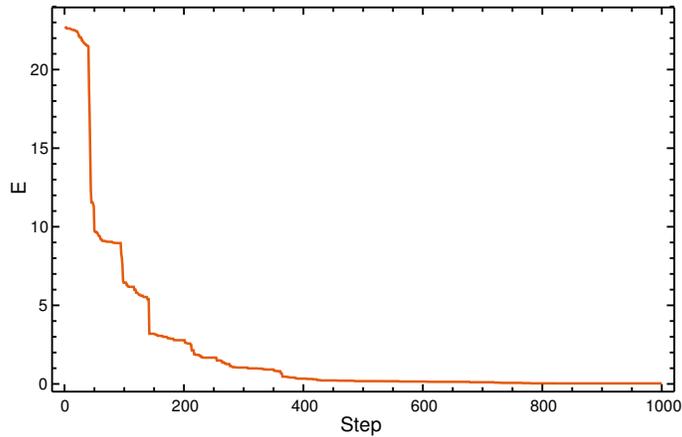}
\caption{ The energy of the system $E$ decreases with the solving steps}
\label{Fig3}
\end{figure}
In order to simulate the imaging processes, we firstly generated the list of positions and
orientations of the NV centers using random numbers and used the list to
generate the simulated CCD image. The positions of each NV centers are discrete
to meet the assumption that each NV center is located at the center of the
CCD pixel, and there are no overlapping NV centers. The imaging plane is perpendicular to the propagation direction of the
light. $\theta $ of each NV
center is evenly distributed between $0$ and $\pi /2$ and $\varphi $ is evenly
distributed between $0$ and $\pi $. The intensity of each NV center in the
image is determined by Eq.(\ref{E1}). A Gaussian filter is applied to the image so that the PSF of each NV centre is a Gaussian matrix.
Then we add a random background noise to
each pixel. Fig.\ref{Fig2}(a)
shows the simulated images of 4 NV centers pumped by different laser polarizations. A total of 12 frames are taken during the imaging process.

Using the method depicted in previous section, the locations and orientations of 4 NV centers are estimated in Fig.\ref{Fig2}(b). The arrows in the estimated images shows the projection of NV orientation vector on the focus plane (xy plane). The directions of the arrows present azimuthal angle $\varphi $ angle, and the lengths of
the arrows are determined by the polar angle $\theta $.
For the PSF with a diameter of $500$ nm, the
resolution of our method can be up to $70$ nm.  From Fig.\ref{Fig3}, we can see that, as the solving steps increase, the
result becomes more similar to the exact answer. It only takes about 400
steps for the energy to become close to $0$, which means the estimated
answer is the same as the real one.

\begin{figure}[tbp]
\centering\includegraphics[width=9cm]{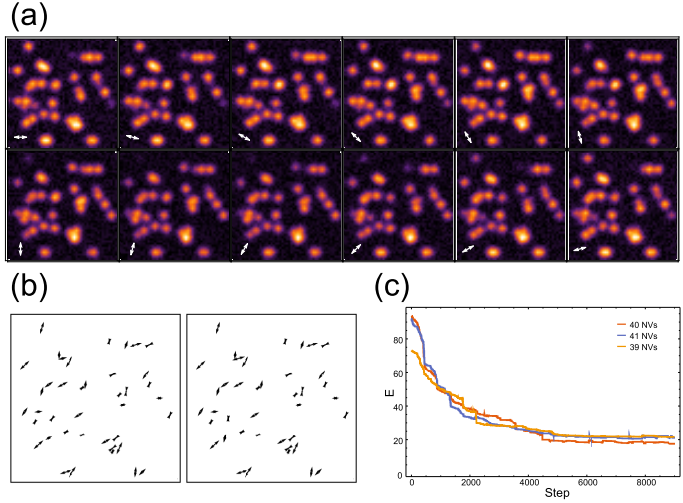}
\caption{(a) Simulated CCD image of 40 NV centers under linear polarized
light with an interval angle of 15 degrees. The background noise is simulated with an intensity about 10\% of a single NV fluorescence intensity. The white arrows show the polarizations of linear polarized laser in each frame.
(b) The distribution of position and orientation of 40 NV centers: left, real positions and orientations; right, estimated result after 10000 steps of solving. (c) Relationship of the
energy of the system with the solving steps using 39,40 and 41 NV centers}
\label{Fig4}
\end{figure}
When it
comes to a system with larger numbers of NV centers, the method also performs well. As shown Fig.\ref{Fig4}, we performed the simulation process to a sample with 40 NV centers. A series of 12 frames are taken with different laser polarizations.
It takes about $4000$ estimation steps for the fit to converge to the final result.
In Fig.\ref{Fig4}(c), we show the convergent energy $E$ with different estimated NV numbers. From the comparison, the number of
NV center should be $40$ with the lowest convergent energy $E$.

For the quantum sensing with NV center, the fluorescence intensity $I(x,y,\varphi ,\theta, \phi )$ changes according to the external field. Taking the external field into the consideration of  system energy $E$, the external field could be measured with high spatial resolution with our method.

\section{Discussing and future work}

Using this polarization demodulation method, we proposed an optical far-field microscopy method that can simultaneously get the position and
orientation of each NV center with high spatial resolution. For the quantum sensing with NV center, the fluorescence intensity of NV center changes according to the external field. The microscopy can be further applied
for the NV center quantum sensing beyond the diffraction limit. This wide field super-resolution microscopy does not need strong laser excitation, which is usually required for STED and CSD microscopy of NV center.  In addition, the fluorescence of NV center is deterministically switched by changing the polarization of pumping laser in our scheme. Comparing with STORM type wide field microscopy, the fluorescence blinking is not need in this method.

The precision of the estimated images is determined by the estimation steps, which is
proportional to the number of NV centers. In the samples discussed above, It
takes about $400$ steps to solve a $4$-NV center system and $4000$ steps to
solve a $40$-NV center system. In each step we need to calculate the total
energy $E$, which is proportional to the solving area. Supposing the number of
NV centers is $N$ and the number of the image's pixels is $S$, then the solving time
is $O(N\times S)$. It takes about $100$ seconds to solve a system with $40$
NV centers and $2500$ pixels, so the solving time is approximately $N\times
S/1000$.

The analyzing process can be  speed up by parallel computing. Large images are divided into several small areas, and then we perform the calculating for each small area. Furthermore, we can also greatly enhance the speed by converting the code to OpenCL or CUDA to use the power of GPU computing.

\end{document}